\documentclass[twocolumn,showpacs,showkeys,preprintnumbers,prd,nofootinbib,superscriptaddress]{revtex4-1}

\usepackage{amsmath}
\usepackage{amsfonts}
\usepackage{amssymb}
\usepackage{graphicx}
\usepackage{color}
\usepackage[]{hyperref}
\usepackage{booktabs}
\usepackage{cleveref}

\Crefname{equation}{Eq.}{Eqs.}
\Crefname{figure}{Fig.}{Figs.}
\Crefname{section}{Sec.}{Secs.}

\usepackage{etoolbox}
\makeatletter
\appto{\appendix}{%
  \@ifstar{\def\theequation@prefix{A.}}%
          {}%
}
\makeatother

\begin{document}

\title{Cosmological viability of a double field unified model from warm inflation}

\author{Rocco D'Agostino}
\email{rocco.dagostino@unina.it}
\affiliation{Scuola Superiore Meridionale, Largo S. Marcellino 10, 80138 Napoli, Italy.}
\affiliation{Istituto Nazionale di Fisica Nucleare (INFN), Sez. di Napoli, Via Cinthia 9, 80126 Napoli, Italy.}

\author{Orlando Luongo}
\email{orlando.luongo@unicam.it}
\affiliation{Dipartimento di Matematica, Universit\`a di Pisa, Largo B. Pontecorvo, 56127 Pisa, Italy.}
\affiliation{Universit\`a di Camerino, Divisione di Fisica, Via Madonna delle carceri, 62032 Camerino, Italy.}
\affiliation{NNLOT, Al-Farabi Kazakh National University, Al-Farabi av. 71, 050040 Almaty, Kazakhstan.}

\begin{abstract}

In this paper, we investigate the cosmological viability of a double scalar field model motivated by warm inflation. To this end, we first set up the theoretical framework in which dark energy, dark matter and inflation are accounted for in a triple unification scheme. We then compute the overall dynamics of the model, analyzing the physical role of coupling parameters. Focussing on the late-time evolution, we test the model against current data. Specifically, using the low-redshift Pantheon Supernovae Ia and Hubble cosmic chronometers measurements, we perform a Bayesian analysis through the Monte Carlo Markov Chains method of integration on the free parameters of the model. We find that the mean values of the free parameters constrained by observations lie within suitable theoretical ranges, and the evolution of the scalar fields provides a good resemblance to the features of the dark sector of the universe. Such behaviour is confirmed by the outcomes of widely adopted selection criteria, suggesting a statistical evidence comparable to that of the standard $\Lambda$CDM cosmology. We finally discuss the presence of large uncertainties over the free parameters of the model and we debate about fine-tuning issues related to the coupling constants.

\end{abstract}


\maketitle

\section{Introduction}

Interpreting the observational evidence that suggests that the universe is currently experiencing an accelerated phase is an outstanding problem for modern cosmology. In particular, there is no consensus about the fluid responsible for the current cosmic speed up \cite{Supernovae,Padmanabhan03,Copeland06}. 
The standard scenario, pictured by the cosmological constant, $\Lambda$, and named the $\Lambda$CDM model, does not provide a fully-satisfactory theoretical interpretation due to the large disagreement between quantum field predictions and observations on the $\Lambda$ magnitude \cite{Weinberg89,Sahni02,Peebles03} and seems to be plagued by alleged cosmic tensions \cite{tensioni,Colgain}. Furthermore, the $\Lambda$CDM paradigm is unable to explain the dark matter nature, which remains not yet understood. The evidence for dark energy and dark matter leads to the so-called \emph{dark sector} 
\cite{Salucci96,Clowe06}. 

A phase of acceleration is not unique throughout the cosmic history. Indeed, at very early stage of universe's evolution, an exponential expansion of space occurred, known as \emph{inflation} \cite{Starobinsky80,Guth81,Linde82}, invoked to explain the tight measurements of the cosmic microwave background depicting a present-day universe that is homogeneous, isotropic and spatially flat up to a high precision \cite{Planck18}.
The inflationary paradigm provides the quantitative predictions for density perturbations that seed  large scale structures. A large amount of data and observational evidences, accumulated during the years, seem to confirm this theory over any other possible alternatives, making inflation the widely-accepted mechanism describing the early-time dynamics  \cite{Planck_inflation}.

As the dark sector nature is unknown, a likely option is to describe dark energy and dark matter under the same standard, making use of dynamical scalar fields  \cite{Sherrer04,Brandenberger19,Bertacca11}. Consequently, one may consider to unify the dark sector with inflation, motivated by the similar dynamical properties between inflation and dark energy and by several attempts toward unified schemes of dark energy-dark matter. This would, in fact, provide a common mechanism responsible for both late and early expansion stages \cite{Peebles99,Capozziello06,Bose09}. 
Thus, as it is possible to identify dark energy, dark matter and inflation with scalar fields, one may think of unifying all these three different phenomena under a single theoretical scheme  \cite{Liddle,Santiago11, Odintsov19}, following a  multifield approach, firstly developed in order to investigate inflationary models within  generalized Einstein theories \cite{Berkin91, Starobinsky01}.

Inspired by this approach, a recent model has been proposed attempting a triple unification based on two scalar fields, one playing the role of inflation and dark matter, while the other one inducing the present acceleration \cite{Sa20,Sa21}. The very early universe, in this case, was described by an alternative form of inflationary mechanism, named \emph{warm inflation} \cite{Berera95,Benetti17}. 
Differently from the standard (cold) inflation, in the warm inflation radiation is produced simultaneously to the inflationary expansion. This implies that the transition between the inflationary phase and the radiation dominated era may occur smoothly without reheating. A further difference with respect to standard inflation concerns the origin of density perturbations. In the warm inflation, they arose from thermal fluctuations rather than vacuum quantum fluctuations. Hence, a great advantage of warm inflation is to identify the dark sector as inflaton field relic, surviving at late times and dominating over matter today.

In this article, we  investigate the cosmological viability of a general two-scalar field model based on warm inflation. We first set up its general formalism and then investigate the late-time evolution of the model, checking its goodness in terms of dynamical variables and the stability of critical points.
In view of the most recent observations, we work out a Markov Chain Monte Carlo (MCMC) analysis to place bounds over the free parameters of the model. 

The paper is structured as follows. After this introduction, in \Cref{sec:theory} we describe the theoretical framework and present the emerging cosmological model. 
In \Cref{sec:dynamics}, we study the dynamics of the model, focussing on its late time evolution through the method of dynamical variables.
Then, in \Cref{sec:observations} we analyze the viability of the model in view of the most recent cosmological observations. Finally,  we discuss our findings, and in \Cref{sec:conclusions} we outline the future perspectives of our work.


\section{A unified cosmological scenario}
\label{sec:theory}

Scalar fields represent  suitable candidates to construct models able to describe early and late-time universe dynamics in lieu of barotropic fluids, playing an important role also in high energy physics \cite{review_particle}. In this respect, cosmological models can be constructed through a single field only, bringing information able to describe the large-scale dynamics. However, the possibility of building multiple scalar field cosmological models is not a mere complication, but it deserves investigation in view of the unification of different cosmic eras \cite{Coley00,Kim05,Ohashi09}. Clearly, those fields may interact with other components, e.g. dark matter, through particular couplings \cite{Amendola00,Holden00,Koivisto05,Gonzalez06}.

We here single out a warm inflation approach by considering an action of the form $S=S_{{\rm HE}}+S_\varphi$ ,
where $S_{{\rm HE}}=\int d^4x \sqrt{-g} \frac{R}{2\kappa^2}$ is the standard Hilbert-Einstein action with $\kappa\equiv \sqrt{8\pi G}$, while $S_\varphi=\int d^4x \sqrt{-g} \mathcal{L}_\varphi$ is a generic scalar field action. 

In particular, motivated by nonstandard kinetic terms and exponential potential emerging in several contexts, among which Kaluza-Klein theories, hybrid metric Palatini theories and $f(R)$ gravity \cite{Rocco_review,DiMarco03,Harko12,Tamanini13}, we write down a double scalar field Lagrangian model:
\begin{align}
\mathcal L_{(\varphi,\,\psi)}=\frac{1}{2}(\nabla\varphi)^2+\frac{1}{2}e^{-\kappa\lambda \varphi}(\nabla\psi)^2+e^{-\kappa\mu\varphi}\ V(\psi)\,,
\label{eq:action}
\end{align}
where $\mu$ and $\lambda$ are dimensionless constants parametrizing the interaction between the scalar fields $\varphi$ and $\psi$, and the nonstandard kinetic term of the field $\psi$, respectively. Hereafter, we refer to the above described double scalar field scenario as \emph{DSF model}.

Under a suitable choice of $V(\psi)$, the DSF model could give rise to a \emph{triple unified cosmological picture}, in which the scalar field $\varphi$ plays the role of dark energy, whereas the role of inflation and dark matter are both played by the scalar field $\psi$. In particular, an appropriate form for the potential can be written in terms of a harmonic oscillator with respect to the scalar field $\psi$ as
\begin{equation}
V(\psi)=V_0+\dfrac{m^2}{2}\psi^2\,,
\label{eq:potential}
\end{equation}
where $V_0$ and $m$ are constants related to the vacuum energy and the mass of the field $\psi$, respectively.
The above potential has been considered in the context of chaotic inflation \cite{Linde02} and investigated as a possible candidate for unified approaches in the string theory \cite{Liddle}. 

The chaotic inflation scenario was introduced to overcome the issues of old and new inflation \cite{Linde83}. These theories assume an initial thermal equilibrium for the universe, which was large and homogeneous enough to survive until the inflationary epoch starts. They can be thought as an incomplete modification of the big bang theory, where inflation represents an intermediate stage of the universe's evolution. Chaotic inflation, instead, can begin even for densities close to the Planck density and in the absence of thermal equilibrium in the very early stages of the universe's evolution. Hence, the advantages of such a scenario lie in the fact that inflation can occur even for potentials with the simplest form $V(\psi)\sim \psi^n$, and that it can provide a simple solution to initial conditions problem (we refer to \cite{Linde90} for more details).
For $|\psi|\gtrsim m_P$, with $m_P$ the Planck mass,  the potential \eqref{eq:potential} drives inflation and, under the condition $m\simeq 10^{-6}m_P$, generates density perturbations as required by observations. Afterwards, the field $\psi$ is subjected to a rapid oscillation over the Hubble time scale, behaving on average as cold dark matter.

In the present scenario, a warm inflation is assumed with a continuous transfer from the inflaton field $\psi$ to a radiation bath, so that the transition to the radiation-dominated epoch occurs smoothly without resorting to a post-inflationary reheating phase. 
The warm-type mechanism differs from the cold one as in the latter the drastic dilution of radiation during inflation needs the presence of a post-inflationary reheating phase to recover the standard evolution of the universe. Throughout the warm inflationary period, instead, the radiation energy density is characterized by a continuous energy transfer between the two scalar fields, which prevents from the dilution of the radiation bath maintaining the universe ``warm". 
The sufficient condition for having warm inflation is the presence of a remarkable energy quantity continuously transferring from the inflaton $\psi$ to radiation, so that one could ignore the energy exchange between the radiation bath and the dark energy field $\varphi$. However, nonvanishing values of the parameters $\lambda$ and $\mu$ in \Cref{eq:action} imply a direct energy exchange between the two fields, which naturally leads to consider energy transfer also between radiation and the field $\varphi$, although this is not strictly required in the warm inflation paradigm. To do that, one can introduce in the equation of motion the dissipating coefficients $\Gamma_\varphi$ and $\Gamma_\psi$.  As dissipative effects increase, radiation starts dominating the universe's evolution and inflation ends. 
The dissipating coefficients are exponentially suppressed soon after inflation and become negligible afterwards.
Immediately after the inflationary epoch, the inflaton field decouples from radiation and oscillates around its potential minimum, showing thus an average behaviour which mimics a non-relativistic fluid with no pressure, giving rise to cold dark matter. In this case, the scalar field $\psi$ does not provide the same evolution of standard baryonic matter due to dependence of its energy density on the scalar field $\varphi$ through the non standard kinetic term in the action. 
Finally, the dark energy behaviour is reproduced by the scalar field $\varphi$, which becomes the predominant component at late times and drives the present observed accelerated expansion of the universe. 
We notice that, for very large $\varphi$, $\mathcal L_{(\varphi,\,\psi)} \approx \mathcal{L}_{\varphi}=\frac{1}{2}(\nabla\varphi)^2$, which corresponds to a free field that cannot drive the universe to accelerate at future times. In other words, large values of $\varphi$ are needful to speed up the universe today, albeit far from $\varphi\rightarrow\infty$.

\section{Dynamics of the model}
\label{sec:dynamics}

In order to study the cosmological dynamics of the two-field model described above, we proceed by varying action \ref{eq:action} with respect to the metric $g_{\mu\nu}$ and the fields $\varphi$, $\psi$. In particular, assuming a flat Friedmann-Lema\^{i}tre-Robertson-Walker metric, one obtains the Friedmann equations
\begin{align}
H^2&=\dfrac{\kappa^2}{3}\left(\rho_r+\dfrac{\dot{\varphi}^2}{2}+\dfrac{{\dot{\psi}}^2}{2}e^{-\kappa\lambda \varphi}+Ve^{-\kappa\mu\varphi}\right) \,, \label{eq:Friedmann1}\\
\dfrac{\ddot{a}}{a} & =-\dfrac{\kappa^2}{3}\left(\rho_r+\dot{\varphi}^2+\dot{\psi}^2e^{-\kappa\lambda\varphi}-Ve^{-\kappa\mu\varphi}\right)\,, \label{eq:Friedmann2}
\end{align}
where $H\equiv\dot{a}/a$ is the Hubble parameter, and $a(t)$ is the cosmic scale factor\footnote{The usual convention is to normalize the scale factor at the present time, \emph{i.e.} $a_0=1$, which corresponds to redshift $z=0$ due to the relation $a\equiv1/(1+z)$.  Throughout the paper, we adopt the notation with a subscript zero to indicate quantities evaluated at the present time.}.

We here consider the equation of state for radiation as $w_r=p_r/\rho_r=\frac{1}{3}$,
and the scalar field evolution given by the Klein-Gordon equations
\begin{align}
&\ddot{\varphi}+3H\dot{\varphi}+\dfrac{\kappa\lambda}{2}\dot{\psi}^2 e^{-\kappa\lambda\varphi}-\kappa \mu Ve^{-\kappa\mu\varphi}=-\Gamma_\varphi\dot{\varphi} \,, \label{eq:KG_fi} \\
&\ddot{\psi}+3H\dot{\psi}-\kappa\lambda\dot{\varphi}\dot{\psi}+\dfrac{\partial V}{\partial \psi}e^{\kappa(\lambda-\mu)\varphi}=-\Gamma_\psi \dot{\psi}e^{\kappa\lambda\varphi}\,, \label{eq:KG_psi}
\end{align}
coupled in the initial inflationary phases with radiation by means of
\begin{equation}
\dot{\rho_r}+4H\rho_r=\Gamma_\varphi \dot{\varphi}^2+\Gamma_\psi \dot{\psi}^2 \,.
\label{eq:KG_rho_r}
\end{equation}
During inflation, the radiation density modifies accordingly to $\Gamma$ friction terms. After the inflationary stage, these terms tend to zero, restoring the common radiation behaviour, namely $\rho_r\sim a^{-4}$.

In order to solve \Cref{eq:Friedmann1,eq:Friedmann2,eq:KG_fi,eq:KG_psi,eq:KG_rho_r} and explore the dynamics of the DSF model, we need to specify the form of the dissipating coefficients $\Gamma_\varphi$ and $\Gamma_\psi$.  
Several forms adopted in the literature usually suppose a functional dependence on the temperature \cite{Berera09,Gil09,Lima19,Rosa19,Gil19}, so that we consider $\Gamma_i\propto T^n$, where $i=(\varphi,\psi)$ and $n\in \mathbb{Z}$. Moreover, we assume that the dissipating coefficients are exponentially suppressed and become negligible soon after inflation. Thus, a suitable form is given by 
\begin{equation}
\Gamma_i=\xi_i \times
\left\{
\begin{aligned}
&T^n\,, \hspace{2cm} T\geq T_f\,, \\
& T^n e^{1-\left(T_f/T\right)^m}\,, \hspace{0.35cm} T\leq T_f\,.
\end{aligned}
\right .
\label{eq:Gamma}
\end{equation} 
Here, $m\in \mathbb{N}_+$, $T_f$ is the radiation temperature at the end of inflation, while $\xi_i>0$ are constants that depend on the microscopic models utilized to obtain the dissipating terms. 
The suppression mechanism plays a key role in the unification picture as it has to guarantee, after inflation, the survival of the field $\psi$, which should acquire enough energy to be able of reproducing the observed dark matter features.
Among the others, it is worth reminding the study done in \cite{Sa20}, where it was shown that, for $n>2$, the dissipating coefficients are naturally suppressed as the temperature becomes lower than a given threshold. This allows to set $m=0$ as the exponential term is actually needless in such a particular scenario. 
Some other possibilities have been explored recently in the literature \cite{Rosa19,Gil19}. These models propose $n\leq 2$ with $m\neq 0$ and are as well characterized by solid observational and theoretical foundations.

\subsection{Late-time evolution} 

We now focus on late-time dynamics of the DSF model, in order to check whether it is able to reproduce the dark energy effects alternatively to the standard cosmological constant scenario. As previously discussed, the dissipation coefficients $\Gamma_\varphi$ and $\Gamma_\psi$ are exponentially suppressed after the inflationary epoch, so that we can set these terms to zero in the following. Also, we can neglect radiation in  \Cref{eq:Friedmann1,eq:Friedmann2}. 
Thus, assuming for the potential the form given in \Cref{eq:potential} and defining the pressure and energy densities of the scalar fields as, respectively,
\begin{align}
p_\varphi &= \dfrac{\dot{\varphi}^2}{2}-V_0e^{-\kappa\mu\varphi} \,, \\
\rho_\varphi&=\dfrac{\dot{\varphi}^2}{2}+V_0e^{-\kappa\mu\varphi} \,,\label{eq:rho_fi}\\
p_\psi & =\dfrac{\dot{\psi}^2}{2}e^{-\kappa\lambda\varphi}-\dfrac{m^2}{2}\psi^2 e^{-\kappa\mu\varphi} \,, \label{eq:pi_psi} \\
\rho_\psi & =\dfrac{\dot{\psi}^2}{2}e^{-\kappa\lambda\varphi}+\dfrac{m^2}{2}\psi^2 e^{-\kappa\mu\varphi} \,,\label{eq:rho_psi}
\end{align}
we can recast \Cref{eq:KG_psi} as  
\begin{equation}
\dot{\rho_\psi}+3H(p_\psi+\rho_\psi)=\dfrac{\kappa}{2}\dot{\varphi}\big(\lambda\dot{\psi}^2e^{-\kappa \lambda \varphi}-\mu m^2\psi^2e^{-\kappa\mu\varphi}\big)\,.
\label{eq:evo rho_psi}
\end{equation}

The analysis can be simplified by taking into account a rapid oscillation of the field $\psi$ around its potential minimum. Such a behaviour is similar to non-relativistic dark matter, so that one can assume, over a period of oscillation, $p_\psi=0$. Thus, combining \Cref{eq:pi_psi,eq:rho_psi}, we find
\begin{equation}
\dot{\psi}^2=\rho_\psi e^{\kappa \lambda\varphi} \,, \quad \psi^2=\dfrac{\rho_\psi}{m^2}e^{\kappa\mu\varphi}\,,
\end{equation}
so that \Cref{eq:evo rho_psi} simplifies to
\begin{equation}\label{contin}
\dot{\rho_\psi}+3H\rho_\psi=Q\,,
\end{equation}
where we obtain the interaction between dark matter and dark energy:
\begin{equation}
Q\equiv \frac{\kappa}{2}(\lambda-\mu)\rho_\psi\dot{\varphi}\,.    
\label{eq:interaction}
\end{equation}
The solution of \Cref{contin} is given by
\begin{equation}
\rho_\psi=\dfrac{\rho_{\psi,0}}{a^3}e^{\frac{\kappa}{2}(\lambda-\mu)(\varphi-\varphi_0)}\,,
\label{eq:matter density}
\end{equation}
where a modification of the standard matter scaling is due to the term $e^{\frac{\kappa}{2}(\lambda-\mu)(\varphi-\varphi_0)}$. In particular, for $\lambda=\mu$, the interaction term $Q$ vanishes,  leading to no energy exchange within the dark sector. In this case, the energy density of the field $\psi$ follows the standard dark matter evolution, $\rho_\psi\sim a^{-3}$, while dark energy evolution is dictated by the term $V_0 e^{-\kappa\mu\varphi}$. On the other hand, for $\lambda\neq \mu$, the dark matter evolution depends on the energy of the field $\varphi$, implying that the amount of dark energy in the matter-dominated era is not negligible and induces an earlier time shift in the transition from the epoch of radiation to the matter-dominated universe. This situation can be avoided by requiring the condition $|\lambda-\mu|\lesssim 1$, in order to be consistent with the predictions of primordial nucleosynthesis \cite{Sa20}.

Our assumptions allow us to rewrite the Friedmann equations as
\begin{align}
H^2&=\dfrac{\kappa^2}{3}\left(\dfrac{\dot{\varphi}^2}{2}+\rho_\psi+V_0e^{-\kappa\mu\varphi}\right) , \label{eq:first_Friedmann} \\
\dot{H}&=-\dfrac{\kappa^2}{2}\left(\dot{\varphi}^2+\rho_\psi\right),
\end{align}
while \Cref{eq:KG_fi} takes the form
\begin{equation}
\ddot{\varphi}+3H\dot{\varphi}-\kappa\mu V_0 e^{-\kappa\mu\varphi}=-\dfrac{\kappa}{2}(\lambda-\mu)\rho_\psi\,.
\end{equation}

Furthermore, introducing the density parameters of cosmic fluids as $\Omega_i\equiv\rho_i/\rho_c$, where $\rho_c=3H^2/\kappa^2$ is the critical density of the universe, from \Cref{eq:rho_fi,eq:rho_psi}, we find
\begin{align}
\Omega_\varphi&=\dfrac{\kappa^2}{3H^2}\left(\dfrac{\dot{\varphi}^2}{2}+V_0e^{-\kappa\mu\varphi}\right),\\
\Omega_\psi&=\dfrac{\kappa^2}{3H^2}\left(\dfrac{\dot{\psi}^2}{2}e^{-\kappa\lambda\varphi}+\dfrac{m^2}{2}\psi^2 e^{-\kappa\mu\varphi}\right),
\end{align}
obeying the constraint $\Omega_\varphi+\Omega_\psi=1$, by virtue of \Cref{eq:first_Friedmann}.

\subsection{Dynamical variables}

To study the cosmological behaviour of the DSF model and its stability, it is convenient to recast the evolution equations by means of dynamical variables. Based on the combination of numerical and analytical techniques,  this method is particularly effective in extracting quantitative information and has been adopted in several cosmological studies, among which \cite{Rocco_teleparallel}.
We thus define the dimensionless quantities
\begin{equation}
x\equiv \dfrac{\kappa \dot{\varphi}}{\sqrt{6}H}\ ,\quad y\equiv \dfrac{\kappa}{\sqrt{3}H}\sqrt{V_0e^{-\kappa\mu\varphi}}\,,
\end{equation}
so that the dynamics is given by the following autonomous system of equations:
\begin{equation}
\left\{
\begin{aligned}
&x'=\dfrac{\sqrt{6}}{2}\mu y^2+\dfrac{3}{2}x(x^2-y^2-1)+\dfrac{\sqrt{6}}{4}(\lambda-\mu)(x^2+y^2-1)\,, \\
&y'=\dfrac{3}{2}y(x^2-y^2+1)-\dfrac{\sqrt{6}}{2}\mu x y\,.
\end{aligned}
\right .
 \label{system}
\end{equation}
Here, the symbol prime indicates a derivative with respect to the number of e-folds, $N\equiv \ln a$. 

Therefore, we can express all the physical quantities inherent to the DSF model in terms of the dynamical variables. In particular, the density parameter of the field $\varphi$ and the dark energy equation of state parameter are given by
\begin{equation}
\Omega_\varphi=x^2+y^2\,,\quad  w_\text{DE}\equiv\dfrac{p_\varphi}{\rho_\varphi}=\dfrac{x^2-y^2}{x^2+y^2}\,.
\end{equation}
Moreover, for the effective equation of state parameter, we find
\begin{equation}
w_\text{eff}\equiv \dfrac{p_\varphi+p_\psi}{\rho_\varphi+\rho_\psi}=x^2-y^2\,,
\end{equation}
while the deceleration parameter takes the form 
\begin{equation}
q=-\dfrac{H'}{H}-1=\dfrac{1}{2}(3x^2-3y^2+1)\,.
\end{equation}

\subsection{Critical points}

Before proceeding to the observational tests of the model, we shall analyze the dynamical system \eqref{system} in terms of critical points and their stability conditions (see \cite{Sa21} for the details). Specifically, it can be shown that there exist five critical points, whose stability can be assessed through the \emph{centre manifold theory} and \emph{Lyapunov's method} \cite{Carr82,Guckenheimer83,Bogoyavlensky85}. 

The first critical point $(I)$ is obtained for $(x,y)=(1,0)$ and is characterized by the eigenvalues 
\begin{subequations}
\begin{align}
\epsilon_1^{(I)}&=3+\frac{\sqrt{6}}{2}(\lambda-\mu)\,, \\
\epsilon_2^{(I)}&=3-\frac{\sqrt{6}}{2}\mu\,.
\end{align}
\end{subequations} 
This implies the presence of an attractor for $\mu\geq\lambda+\sqrt{6}$, a saddle point for $\sqrt{6}\leq\mu<\lambda+\sqrt{6}$ and a repeller for $\mu<\sqrt{6}$.
Moreover, in this case, we have a stiff-matter fluid $(w_\text{eff}=1)$ in a universe completely dominated by the kinetic term of $\varphi$ $(\Omega_\varphi=1)$.

The second critical point $(II)$ is for  $(x,y)=(1,0)$ and the corresponding eigenvalues are
\begin{subequations}
\begin{align}
\epsilon_1^{(II)}&=3+\frac{\sqrt{6}}{2}\mu\,, \\
\epsilon_2^{(II)}&=3-\frac{\sqrt{6}}{2}(\lambda-\mu)\,.
\end{align}
\end{subequations} 
The critical point is an attractor for $\mu\leq-\sqrt{6}$, a saddle point for $-\sqrt{6}<\mu\leq\lambda-\sqrt{6}$, and a repeller for $\mu> \lambda-\sqrt{6}$. Physically, these results describe the same situation of the critical point $I$, in which the dynamics of the universe is entirely unaffected by the presence of the scalar field $\psi$ $(\Omega_\psi=0)$, whose contribution become negligible around the critical point.

A third critical point $(III)$ is found for $x=(\mu-\lambda)/\sqrt{6}$ and $y=0$. It exists if $|\lambda-\mu|\leq\sqrt{6}$ and the relative eigenvalues are
\begin{subequations}
\begin{align}
\epsilon_1^{(III)}&=\frac{3}{2}+\dfrac{\lambda^2-\mu^2}{4}\,, \\
\epsilon_2^{(III)}&=-\frac{3}{2}+\dfrac{(\lambda-\mu)^2}{4}\,.
\end{align}
\end{subequations} 
Such a point is an attractor for $\sqrt{\lambda^2+6}\leq\mu\leq \lambda+\sqrt{6}$, and a saddle point for $\lambda-\sqrt{6}\leq \mu <\sqrt{\lambda^2+6}$. 
In this case, the cosmic dynamics depends on the particular values of the coefficients $\lambda$ and $\mu$, since $\Omega_\varphi=w_\text{eff}=(\lambda-\mu)^2/6$.
Therefore, if $\lambda=\mu$, we have $\Omega_\psi=1$, so that the universe is dominated by dark matter. On the other hand, $\mu=\lambda\pm\sqrt{6}$ reproduces a stiff-matter behaviour with $\Omega_\varphi=1$, while intermediate cases correspond to relative domination of one field over the other,  with $0\leq w_\text{eff} \leq 1$. It is worth to note that, in the case of a saddle point solution in which the field $\psi$ dominates, one has a period of matter domination necessary for structure formation.

The fourth critical point $(IV)$ is obtained for $x=\mu/\sqrt{6}$ and $y=\sqrt{1-\mu^2/6}$, and exists for $|\mu|\leq \sqrt{6}$. It is characterized by the eigenvalues
\begin{subequations}
\begin{align}
\epsilon_1^{(IV)}&=-3+\frac{\mu}{2}(\lambda+\mu)\,, \\
\epsilon_2^{(IV)}&=-3+\frac{\mu^2}{2} \,.
\end{align}
\end{subequations} 
One thus finds an attractor solution for $-\sqrt{6}\leq\mu\leq(\sqrt{\lambda^2+24}-\lambda)/2$, and a saddle point for $(\sqrt{\lambda^2+24}-\lambda)/2<\mu\leq \sqrt{6}$.
Here, the field $\varphi$ dominates the universe's evolution $(\Omega_\varphi=1)$, while $-1\leq w_\text{eff}\leq 1$, where the lower bound is obtained for $\mu=0$ and provides an accelerated expansion typical of the cosmological constant. On the other hand, $w_\text{eff}=1$ holds if $\mu=\pm\sqrt{6}$, describing a stiff-matter fluid dominated by the kinetic term of the field $\varphi$. Moreover, an interesting situation occurs for $|\mu|<\sqrt{2}$, which corresponds to an accelerated expansion with $w_\text{eff}<-1/3$.

Finally, the fifth critical point $(V)$ corresponds to $x=\sqrt{6}/(\lambda+\mu)$ and $y=\sqrt{\lambda^2-\mu^2+6}/(\lambda+\mu)$. It exists for $(-\lambda+\sqrt{\lambda^2+24})/2\leq \mu \leq \sqrt{\lambda^2+6}$, and its eigenvalues are
\begin{equation}
\epsilon_{1,2}^{(V)}=-\dfrac{3\lambda}{2(\lambda+\mu)}(1\pm\sqrt{1+\mathcal{C}})\,,
\end{equation}
where $\mathcal{C}\equiv 2(\mu^2+\mu\lambda-6)(\mu^2-\lambda^2-6)/(3\lambda^2)$. In this case, within its domain of existence, the critical point represents an attractor solution. Depending on the $\lambda$ and $\mu$ values, the cosmic evolution can be dominated by either fields, as $\Omega_\varphi=(\lambda^2-\mu^2+12)/(\lambda+\mu)^2$ and $w_\text{eff}=(\mu-\lambda)/(\mu+\lambda)$. In particular, taking into account the lower bound of the domain of existence, we have a universe with  $\Omega_\varphi=1$ and $w_\text{eff}=-1$ for $\lambda\rightarrow1$. Conversely, for the upper bound of the domain of existence of the critical point, we have $\Omega_\psi=1$ and $w_\text{eff}=1$ as $\lambda\rightarrow\infty$. Here, the interesting case, corresponding to an accelerated expansion with $w_\text{eff}<-1/3$, occurs for $\mu<\lambda/2$ and $\mu>(-\lambda+\sqrt{\lambda^2+24})/2$. Moreover, we note that the ratio between the densities of the fields is given as
\begin{equation}
\dfrac{\Omega_\varphi}{\Omega_\psi}=\dfrac{\lambda^2-\mu^2+12}{2(\mu^2+\mu\lambda-6)}\,,
\end{equation}
which represents the so-called \emph{scaling solution} \cite{Wetterich88}, frequently used to heal the coincidence problem. Hence, for the particular conditions $\mu<\lambda/2$ and $\mu>(-\lambda+2\sqrt{\lambda^2+18})/3$ (with $\lambda>4\sqrt{2}$), we would have $\Omega_\psi>\Omega_\varphi$ and, thus, an accelerated expansion driven by the field $\varphi$ together with the field $\xi$ behaving as dark energy. However, as discussed earlier, the condition $|\lambda-\mu|\lesssim 1$ needed for the consistency with big-bang nucleosynthesis requires to consider only the values of $\lambda$ and $\mu$ for which dark energy is not described by the field $\psi$.

\section{Observational viability}
\label{sec:observations}

In this section, we present the results of our observational tests on the DSF model. Our study is performed through a direct comparison with low-redshift data currently available from cosmological surveys. 

We specifically focus on data that do not rely on any fiducial cosmology, in order to avoid possible bias towards preferred physical outcomes. Due to this reason, for example, we do not take into account the most common baryon acoustic oscillations (BAO) measurements, which effectively depend on the $\Lambda$CDM model, assumed as a fiducial cosmological basis
Instead, we use the type Ia Supernovae (SN) data provided by the Pantheon catalogue \cite{Scolnic18}, which has been used to construct model-independent measurements of the inverse reduced Hubble parameter, namely $E^{-1}(z)\equiv H(z)/H_0$ \cite{Riess18}. These measurements, indeed, only rely on the hypothesis of a flat universe consistently with what assumed here. 
Moreover, we also consider the cosmic chronometers (CC) data acquired by means of the differential age method \cite{Jimenez02}. Based on spectroscopic estimates of the age of nearby red galaxies, such data provide the Hubble parameter as $H(z)=-(1+z)^{-1}dz/dt$.
Therefore, SN and CC data represent reliable and robust model-independent measurements ensuring unbiased observational tests \cite{D'Agostino19}.

\subsection{Markov Chain Monte Carlo analysis}

We numerically solved the autonomous system (\ref{system}) with suitable initial conditions set in the radiation-dominated era. These conditions have been chosen to be consistent with the most recent measurements \cite{Planck18} suggesting that dark energy represents about $70\%$ of the total density of the universe today, and that the transition from matter-dominated era to dark energy-dominated era happened at $N\approx-5$.
Thus, we were able to find the Hubble expansion rate from integrating the following differential equation:
\begin{equation}
\dfrac{H'}{H}=-\dfrac{3}{2}(1+x^2-y^2)\,,
\end{equation}
with initial condition $H(N=0)=H_0$, where $H_0$ is the Hubble constant.

To compare the model with observations, we then applied the MCMC integration method through a Bayesian analysis performed on the combination of SN and CC measurements.
To this end, we built the likelihood function of the SN data as 
\begin{equation}
L_\text{SN}\propto \exp\left\{-\dfrac{1}{2}\mathbf{v}^\text{T}\mathbf{C}_\text{SN}^{-1}\,\mathbf{v}\right\},
\end{equation}
where $\mathbf{C}_\text{SN}$ is the covariance matrix accounting for the correlation among the 6 measurements provided in \cite{Riess18}, while the differences between the theoretical and observed values are encoded in the vector $\mathbf{v}$, such that $v_i\equiv E^{-1}_{obs,i}-E^{-1}_{th}(z_i)$.

Similarly, for the CC data we constructed the likelihood function for the 31 measurements collected in \cite{Capozziello18} as
\begin{equation}
L_\text{CC}\propto\exp \left\{-\dfrac{1}{2}\displaystyle{\sum_{i=1}^{31}}\left[\dfrac{H_{obs,i}-H_{th}(z_i)}{\sigma_i}\right]^2\right\},
\end{equation}
where $\sigma_i$ are the measurement uncertainties. 

Therefore, the constraints on the model were obtained using the joint likelihood of the SN+CC data, which is given by
\begin{equation}
L_\text{joint}=L_\text{SN} \times L_\text{CC}\,.
\end{equation}

\begin{figure}
\begin{center}
\includegraphics[width=3.2 in]{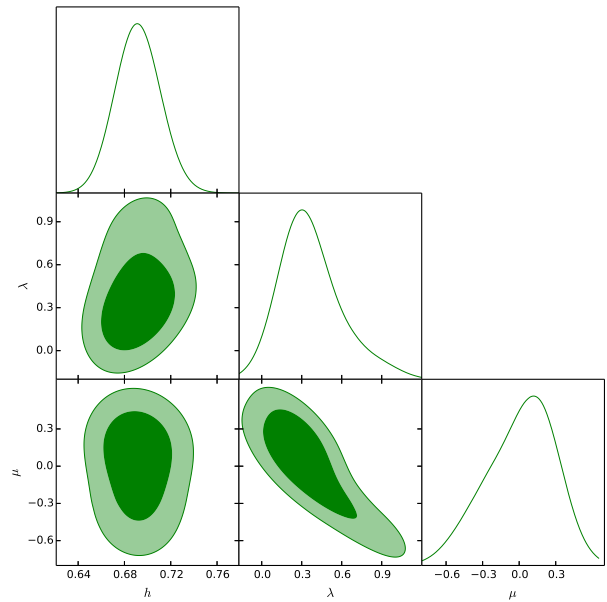}
\caption{1$\sigma$ and 2$\sigma$ confidence regions, with posterior distributions, for the parameters of the DSF model as a result of the MCMC analysis on the Pantheon+CC measurements.}
\label{fig:contours}
\end{center}
\end{figure}

Assuming uniform priors for the free parameters of the DSF model, the results of our MCMC analysis at  $68\%$ confidence level provide\footnote{In our numerical study, we consider the dimensionless hubble parameter $h\equiv H_0/(100$ km s$^{-1}$ Mpc$^{-1})$.}.
\begin{subequations}
\begin{align}
&h=0.692\pm 0.018\ , \label{eq:H0}\\
&\lambda=0.36^{+0.18}_{-0.26}\ ,  \\
&\mu=0.01^{+0.34}_{-0.24} \ .
\end{align}
\end{subequations}
We show in \Cref{fig:contours} the marginalized two-dimensional contour regions at 68\% and 95\% confidence levels.

\subsection{Cosmological consequences}

\begin{figure}
\begin{center}
\includegraphics[width=3.2 in]{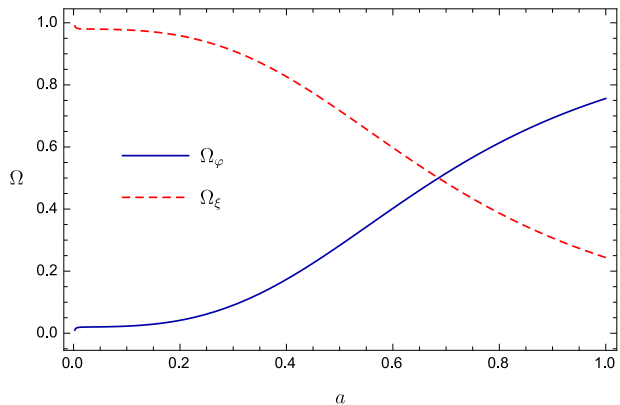}
\caption{Cosmic evolution of the density parameters of the scalar fields for the DSF model. The curves are calculated for the mean values obtained from the MCMC analysis.}
\label{fig:densities}
\end{center}
\end{figure}

\begin{figure}
\begin{center}
\includegraphics[width=3.2 in]{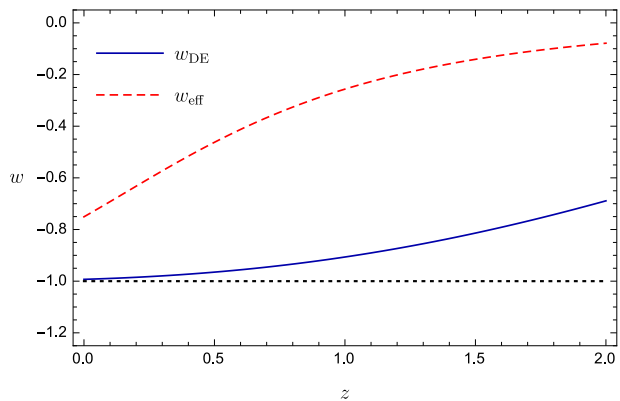}
\caption{Evolution of the effective and the dark energy equation of state parameters, as a function of the redshift, for the double field unification model corresponding to the mean results of the MCMC analysis. The dotted line represents the cosmological constant case.}
\label{fig:w}
\end{center}
\end{figure}

\begin{figure}[h!]
\begin{center}
\includegraphics[width=3.2 in]{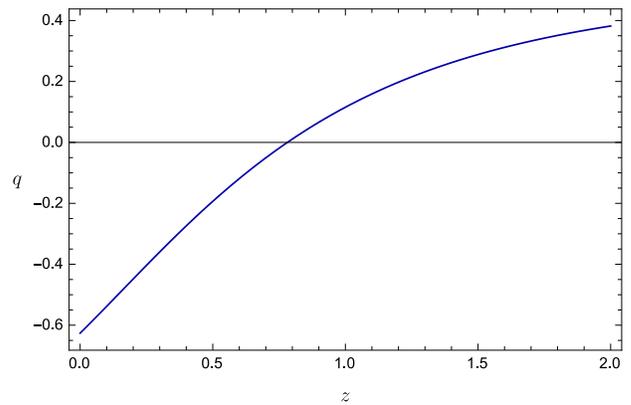}
\caption{Redshift behaviour of the deceleration parameter for the double field unification model assuming the mean values of the MCMC results. The solid black line marks the divide between a decelerating universe $(q>0)$ and an accelerating universe $(q<0)$.}
\label{fig:q}
\end{center}
\end{figure}

Using the results presented above, we were able to place bounds on the density parameter of the field $\psi$.   In particular, we obtained 
\begin{equation}
\Omega_{\psi,0}=0.244\pm{0.056}\,,  
\end{equation}
which is perfectly consistent with the estimated amount of dark matter in the present universe \cite{Planck18}.
\Cref{fig:densities} shows the evolution of the scalar fields density parameters as a function of the scale factor. We notice the ability of the DSF model to reproduce the correct sequence of the cosmic evolution: 

\begin{itemize}
    \item[-] early eras are dominated by the scalar field $\psi$, being responsible for structures formation and behaving like cold dark matter;
    \item[-]  the scalar field $\varphi$ dominates the late evolution of the universe, mimicking dark energy and driving the accelerated expansion era.
\end{itemize}

We also display in \Cref{fig:w} the late-time evolution of the dark energy equation of state parameter and the effective equation of state parameter. As one can see, the scalar field $\varphi$ reproduces the features of  $\Lambda$, approaching the value $-1$ at recent times. 

Furthermore, the behaviour of the deceleration parameter at recent times is shown in \Cref{fig:q}. One can thus calculate the epoch of transition between the of matter-dominated and the dark energy-dominated eras. In the case of the DSF model, the mean values of the MCMC analysis predict that such a transition occurs at $a \simeq 0.56\,.$

\subsection{Statistical significance}

Our results have been compared to the predictions of the standard $\Lambda$CDM model. In particular, from the combined SN+CC data analysis, we found \cite{D'Agostino20}
\begin{subequations}
\begin{align}
&h=0.692 \pm 0.019\,, \\
&\Omega_{m,0}=0.296^{+0.026}_{-0.029}\,.
\end{align}
\end{subequations}
It is worth to remark that the value of the Hubble constant emerging from the DSF model (c.f. \Cref{eq:H0}) is fully consistent with the outcome predicted by the $\Lambda$CDM scenario for the specific choice of data used in the present study.

The evidence for the DSF model against standard cosmology has been measured by making use of statistical estimators, such as information criteria. To this end, we adopted the Akaike information criterion (AIC) \cite{Akaike74}, which corrects the maximum likelihood value by the numbers of free parameters of a model, and the even more accurate DIC criterion \cite{Kunz06}, taking into account the effective number of parameters constrained by the data considered in the analysis. 
Thus, one can select the best theoretical scenario that coincides with the lower values of information criteria. This is provided by the difference $\Delta$ between the selection criteria values of the models under comparison. Usually, we can distinguish among the following cases: $\Delta<0$ indicates statistical preference for the specific model over the reference model; if $0<\Delta< 2$, the model is weakly disfavoured with respect to the reference scenario; $2<\Delta<6$ suggests mild evidence against the model; for $\Delta>6$, the model is strongly disfavoured compared to the reference model. 

In our study, we tested the DSF model against the reference $\Lambda$CDM scenario, obtaining the following results:
\begin{equation}
\Delta\text{AIC}=1.02\ , \quad \Delta\text{DIC}= -0.61\ .
\end{equation}
These values show the excellent statistical performance of the DSF model at late times.  

\subsection{Theoretical discussion}

Despite the ability of the DSF model to accommodate for low-redshift cosmic data, its cosmological viability remains, however, plagued by the large uncertainties over the free parameters of the model. Also, the constraint $|\lambda-\mu|\lesssim1$ imposed by the observations of primordial nucleosynthesis, although consistent with the results of our numerical analysis, requires a fine tuning on the coupling constants that cannot be ignored.

In addition, to clarify the physical meaning  of $\lambda$ and $\mu$, it is possible to work out the limit of very small $\varphi$. Even if  this approximation does not address experimental evidence, as stressed above, it is interesting to notice that, neglecting third and higher-order field contributions and noticing that linear field terms into the Lagrangian provide only a shift in $\varphi$, one would have $\mathcal L_{(\varphi,\psi)}^{(\varphi\ll1)}=\frac{1}{2}(\nabla\varphi)^2+\frac{1}{2}(\nabla\psi)^2-\frac{\kappa^2\mu^2|V_0|}{2}\varphi^2+\frac{m^2}{2}\psi^2$. Here, a further drawback of the model arises: the bare mass of the $\psi$ constituent turns out to be ill-defined. Indeed, taking negative $V_0$, it is possible to define a physical constituent for $\varphi$, i.e. $
m_{bare,\,\varphi}^2=\kappa^2\mu^2|V_0|$. However, the opposite happens for $\psi$, as the \emph{psion} particle would have $ 
m_{bare,\,\psi}^2=-m^2$, which clearly leads to a tachyon nature of the corresponding inflaton and dark matter species\footnote{A recent study showing how inflaton field can provide a quasi-particle contribution to dark matter has been developed in  \cite{barioni}.} as $\varphi\ll1$. Consequently,  we are forced to take large values of $\varphi$ only. This request  guarantees  inflaton and dark matter to be physical particles, but implies a further tight constraint over the underlying model. 
Moreover, for very small $\varphi$, the two fields do not interact with each other and, thus, the interaction between inflaton and dark energy can occur only as $\varphi$ increases. 

The above theoretical considerations may imply restrictions on the model parameters. However, the  infrared limit for $\varphi$, namely $\varphi\ll1$, does not put severe constraints on the fitting parameters, but rather suggests the energy regime to handle for $\varphi$ throughout the entire universe's evolution. Consequently, we cannot use those theoretical bounds as priors for our fits, since the fields are not upper-bounded. As emerging from our analysis, the nucleosynthesis requirements are satisfied, guaranteeing $|\lambda-\mu|\lesssim1$, which turns out to be the unique effective constraint to address. Nevertheless, another relevant aspect is related to $m_{bare,\varphi}$ that depends on $V_0$. The offset $V_0$ is conjectured to fix early vacuum energy  \cite{LM} and, in principle, could be jeopardized by the cosmological constant problem \cite{marty}. Thus, it appears clear that  more sophisticated versions of the employed potential could modify the bare masses and can fix the above restrictions on the field values. This may help in developing a more accurate ultraviolet theory that works well even at the infrared energy regime.

\section{Conclusions and perspectives}
\label{sec:conclusions}

We here investigated the dynamical and late-time properties of a cosmological model where two scalar fields produce a triple unified paradigm able to unify the properties of inflation, dark energy and dark matter. 
The theoretical foundation of the DSF model is the warm inflation, which suggests a non-trivial coupling between two fields in the gravitational action. 

In the present study, we first outlined the evolution properties of the DSF model, and then we investigated the late-time behaviour in terms of dynamical variables. We thus analyzed the critical points and their stability conditions, providing a physical interpretation of the obtained results.  
Afterwards, we studied the observational viability of the scenario under consideration through a comparison with low-redshift data.  To do so, we worked out a MCMC analysis, obtaining constraints over the free parameters of the model. We thus compared our findings with those from the standard $\Lambda$CDM cosmology, testing the statistical evidence of the DSF model by means of AIC and DIC information criteria. 

Our results show that the DSF model is able to reproduce the correct sequence of cosmological epochs, and the evolution of the scalar fields at recent times mimic the features of the dark sector in agreement with the latest observational evidences. Moreover, statistical criteria certify that the DSF model performs quite similarly to the $\Lambda$CDM paradigm, which is considered the favourite benchmark for describing cosmology. 
 
However, even though the DSF model is experimentally suitable, the underlying free parameters are not well constrained, limiting the numerical analysis itself. 
Also, the offset $V_0$, intimately related to the cosmological constant problem, is left unconstrained. 
The criterion $|\lambda-\mu|\lesssim1$ is observationally fulfilled, but poses fine-tuning issues over the ranges of the underlying free parameters. 

For these reasons, it appears evident that the complex couplings proper of the DSF Lagrangian may lead to non-conclusive results and deserve particular attention in order to establish the goodness of the model. In fact, the poorly constrained coupling coefficients could be due to the choice of data sets here involved. Therefore, further investigations would be helpful to assess the model in more detail at early times. 
In particular, cosmic microwave background (CMB) and large scale structure data will likely place severe constraints on the model.
In this respect, we want to emphasize that several cosmological models having a scalar field with an exponential potential such that the interaction term is of the form $Q\propto \rho \dot{\varphi}$ as in \Cref{eq:interaction} have been investigated at the perturbation level (see \emph{e.g.} \cite{Billyard00,Valentini02,Gumjudpai05,Bohmer08,Tzanni14}), and they proved to be in agreement with both CMB anisotropies and observations of structure formation. Such an analysis will thus be the main subject of upcoming works as it may definitely confirm the viability of the DSF model.

\begin{acknowledgements}

R.D. would like to thank Paulo M. S\'a for useful discussions. The authors are grateful to the anonymous referee for  valuable comments and suggestions that allowed to improve the quality of the manuscript.
R.D. acknowledges INFN (iniziativa specifica QGSKY) for support.
O.L. acknowledges funds from the Ministry of Education and Science of the Republic of Kazakhstan, Grant: IRN AP08052311.

\end{acknowledgements}

\end{document}